\documentstyle[twoside,epsfig]{article}

\catcode`\@=11
\long\def\@makefntext#1{
\protect\noindent \hbox to 3.2pt {\hskip-.9pt
$^{{\eightrm\@thefnmark}}$\hfil}#1\hfill}       

\def\thefootnote{\fnsymbol{footnote}}
\def\@makefnmark{\hbox to 0pt{$^{\@thefnmark}$\hss}}    

\def\ps@myheadings{\let\@mkboth\@gobbletwo
\def\@oddhead{\hbox{}
\rightmark\hfil\eightrm\thepage}
\def\@oddfoot{}\def\@evenhead{\eightrm\thepage\hfil
\leftmark\hbox{}}\def\@evenfoot{}
\def\sectionmark##1{}\def\subsectionmark##1{}}

\oddsidemargin=\evensidemargin
\renewcommand{\thefootnote}{\fnsymbol{footnote}}

\newcounter{sectionc}\newcounter{subsectionc}\newcounter{subsubsectionc}
\renewcommand{\section}[1] {\vspace{12pt}\addtocounter{sectionc}{1}
\setcounter{subsectionc}{0}\setcounter{subsubsectionc}{0}\noindent
    {\tenbf\thesectionc. #1}\par\vspace{5pt}}
\renewcommand{\subsection}[1] {\vspace{12pt}\addtocounter{subsectionc}{1}
    \setcounter{subsubsectionc}{0}\noindent
    {\bf\thesectionc.\thesubsectionc. {\kern1pt \bfit #1}}\par\vspace{5pt}}
\renewcommand{\subsubsection}[1] {\vspace{12pt}\addtocounter{subsubsectionc}{1}
    \noindent{\tenrm\thesectionc.\thesubsectionc.\thesubsubsectionc.
    {\kern1pt \tenit #1}}\par\vspace{5pt}}
\newcommand{\nonumsection}[1] {\vspace{12pt}\noindent{\tenbf #1}
    \par\vspace{5pt}}

\newcounter{appendixc}
\newcounter{subappendixc}[appendixc]
\newcounter{subsubappendixc}[subappendixc]
\renewcommand{\thesubappendixc}{\Alph{appendixc}.\arabic{subappendixc}}
\renewcommand{\thesubsubappendixc}
    {\Alph{appendixc}.\arabic{subappendixc}.\arabic{subsubappendixc}}

\renewcommand{\appendix}[1] {\vspace{12pt}
        \refstepcounter{appendixc}
        \setcounter{figure}{0}
        \setcounter{table}{0}
        \setcounter{lemma}{0}
        \setcounter{theorem}{0}
        \setcounter{corollary}{0}
        \setcounter{definition}{0}
        \setcounter{equation}{0}
        \renewcommand{\thefigure}{\Alph{appendixc}.\arabic{figure}}
        \renewcommand{\thetable}{\Alph{appendixc}.\arabic{table}}
        \renewcommand{\theappendixc}{\Alph{appendixc}}
        \renewcommand{\thelemma}{\Alph{appendixc}.\arabic{lemma}}
        \renewcommand{\thetheorem}{\Alph{appendixc}.\arabic{theorem}}
        \renewcommand{\thedefinition}{\Alph{appendixc}.\arabic{definition}}
        \renewcommand{\thecorollary}{\Alph{appendixc}.\arabic{corollary}}
        \noindent{\tenbf Appendix \theappendixc #1}\par\vspace{5pt}}
\newcommand{\subappendix}[1] {\vspace{12pt}
        \refstepcounter{subappendixc}
        \noindent{\bf Appendix \thesubappendixc. {\kern1pt \bfit #1}}
    \par\vspace{5pt}}
\newcommand{\subsubappendix}[1] {\vspace{12pt}
        \refstepcounter{subsubappendixc}
        \noindent{\rm Appendix \thesubsubappendixc. {\kern1pt \tenit #1}}
    \par\vspace{5pt}}

\newcommand{\textlineskip}{\baselineskip=13pt}
\newcommand{\smalllineskip}{\baselineskip=10pt}

\def\eightcirc{
\begin{picture}(0,0)
\put(4.4,1.8){\circle{6.5}}
\end{picture}}
\def\eightcopyright{\eightcirc\kern2.7pt\hbox{\eightrm c}}

\newcommand{\copyrightheading}[1]
    {
      }

\newcommand{\publisher}[2]{{\begin{center}\footnotesize\smalllineskip
    Received #1\\
    Revised #2
    \end{center}
    }}

\def\abstracts#1#2#3{{
    \centering{\begin{minipage}{4.5in}\footnotesize\baselineskip=10pt
    \parindent=0pt #1\par
    \parindent=15pt #2\par
    \parindent=15pt #3
    \end{minipage}}\par}}

\def\keywords#1{{
    \centering{\begin{minipage}{4.5in}\footnotesize\baselineskip=10pt
    {\footnotesize\it Keywords}\/: #1
    \end{minipage}}\par}}

\newcommand{\bibit}{\nineit}
\newcommand{\bibbf}{\ninebf}
\renewenvironment{thebibliography}[1]
        {\frenchspacing
     \ninerm\baselineskip=11pt
         \begin{list}{\arabic{enumi}.}
        {\usecounter{enumi}\setlength{\parsep}{0pt}
     \setlength{\leftmargin 12.7pt}{\rightmargin 0pt} 
         \setlength{\itemsep}{0pt} \settowidth
    {\labelwidth}{#1.}\sloppy}}{\end{list}}

\newcounter{itemlistc}
\newcounter{romanlistc}
\newcounter{alphlistc}
\newcounter{arabiclistc}

\newcommand{\fcaption}[1]{
        \refstepcounter{figure}
        \setbox\@tempboxa = \hbox{\footnotesize Fig.~\thefigure. #1}
        \ifdim \wd\@tempboxa > 5in
           {\begin{center}
        \parbox{5in}{\footnotesize\smalllineskip Fig.~\thefigure. #1}
            \end{center}}
        \else
             {\begin{center}
             {\footnotesize Fig.~\thefigure. #1}
              \end{center}}
        \fi}

\newcommand{\tcaption}[1]{
        \refstepcounter{table}
        \setbox\@tempboxa = \hbox{\footnotesize Table~\thetable. #1}
        \ifdim \wd\@tempboxa > 5in
           {\begin{center}
        \parbox{5in}{\footnotesize\smalllineskip Table~\thetable. #1}
            \end{center}}
        \else
             {\begin{center}
             {\footnotesize Table~\thetable. #1}
              \end{center}}
        \fi}

\def\@citex[#1]#2{\if@filesw\immediate\write\@auxout
    {\string\citation{#2}}\fi
\def\@citea{}\@cite{\@for\@citeb:=#2\do
    {\@citea\def\@citea{,}\@ifundefined
    {b@\@citeb}{{\bf ?}\@warning
    {Citation `\@citeb' on page \thepage \space undefined}}
    {\csname b@\@citeb\endcsname}}}{#1}}

\newif\if@cghi
\def\cite{\@cghitrue\@ifnextchar [{\@tempswatrue
    \@citex}{\@tempswafalse\@citex[]}}
\def\citelow{\@cghifalse\@ifnextchar [{\@tempswatrue
    \@citex}{\@tempswafalse\@citex[]}}
\def\@cite#1#2{{$\null^{#1}$\if@tempswa\typeout
    {IJCGA warning: optional citation argument
    ignored: `#2'} \fi}}

\def\pmb#1{\setbox0=\hbox{#1}
    \kern-.025em\copy0\kern-\wd0
    \kern.05em\copy0\kern-\wd0
    \kern-.025em\raise.0433em\box0}

\def\fnt#1#2{\footnotetext{\kern-.3em
    {$^{\mbox{\scriptsize #1}}$}{#2}}}

\def\ps@myheadings{%
    \let\@oddfoot\@empty\let\@evenfoot\@empty
    \def\@evenhead{\slshape\leftmark\hfil}
    \def\@oddhead{\hfil{\slshape\rightmark}}
    \let\@mkboth\@gobbletwo
    \let\sectionmark\@gobble
    \let\subsectionmark\@gobble
    }
\font\tenrm=cmr10
\font\tenit=cmti10
\font\tenbf=cmbx10
\font\bfit=cmbxti10 at 10pt
\font\ninerm=cmr9
\font\nineit=cmti9
\font\ninebf=cmbx9
\font\eightrm=cmr8

\def\qed{\hbox{${\vcenter{\vbox{            
   \hrule height 0.4pt\hbox{\vrule width 0.4pt height 6pt
   \kern5pt\vrule width 0.4pt}\hrule height 0.4pt}}}$}}

\renewcommand{\thefootnote}{\fnsymbol{footnote}}    

\def\bsc{{\sc a\kern-6.4pt\sc a\kern-6.4pt\sc a}}   
\def\bflatex{\bf L\kern-.30em\raise.3ex\hbox{\bsc}\kern-.14em
T\kern-.1667em\lower.7ex\hbox{E}\kern-.125em X}

\newcommand{\lb}{{<}}
\newcommand{\rb}{{>}}
\begin{document}
\setlength{\textheight}{7.7truein}  

\thispagestyle{empty}

\markboth{\protect{\footnotesize\it Equilibrium
times}}{\protect{\footnotesize\it Equilibrium times}}

\normalsize\textlineskip

\setcounter{page}{1}

\copyrightheading{}         

\vspace*{0.88truein}

\centerline{\bf EQUILIBRIUM TIMES FOR THE MULTICANONICAL METHOD}
\vspace*{0.035truein} \vspace*{0.37truein}
\centerline{\footnotesize M. L. Guerra and J.D. Mu\~noz} \baselineskip=12pt
\centerline{\footnotesize\it Dpto. de F\'{\i}sica, Univ. Nacional
de Colombia} \baselineskip=10pt \centerline{\footnotesize\it
Bogot\'{a} D.C, Colombia} \centerline{\footnotesize\it E-mail:
margue@linuxmail.org, jdmunozc@unal.edu.co}

\publisher{(received date)}{(revised date)}

\vspace*{0.25truein} \abstracts{This work measures the time to
equilibrium for the multicanonical method on the 2D-Ising system
by using a new criterion, proposed here, to find the time to
equilibrium, $t_{\rm eq}$, of any sampling procedure based on a
Markov process. Our new procedure gives the same results that the
usual one, based on the magnetization, for the canonical
Metropolis sampling on a 2D-Ising model at several temperatures.
For the multicanonical method we found a power-law relationship
with the system size, $L$, of $t_{\rm eq}$$=$$0.27(15)
L^{2.80(13)}$ and with the number of energy levels to explore,
$k_{\rm E}$, of $t_{\rm eq}$$=$$0.7(13) k_{\rm E}^{1.40(11)}$, in
perfect agreement with the result just above. In addition, a kind
of {\it critical slowing
  down} was observed around the critical energy. Our new procedure is
completely general, and can be applied to any sampling method
based on Markov processes.}{}{}

\vspace*{5pt} \keywords{Multicanonical method, time to
equilibrium, Markov processes, flat-histogram algorithms}


\vspace*{1pt}\textlineskip \setcounter{footnote}{0}
\renewcommand{\thefootnote}{\alph{footnote}}
\section{Introduction}

The multicanonical method, first proposed by B. Berg and T.
Neuhaus \cite{berg,berg0} and also known as entropic sampling
\cite{lee}, is a powerful method to explore the phase space of
many physical systems on broad energy ranges. It has been
sucessfully employed, not only to the study of both first and
second order phase transitions in statistical mechanics
\cite{berg1,berg2,bergBilloireJanke2002}, but also to a plenty of
many other areas, from optimization problems in computational
sciences \cite{LimaMenezes} to protein folding
\cite{PengHansmann2003,ArkinCelik,HansmannOkamoto} and even
covering rules for quasicrystals \cite{ReichertGahler} (for a
review on the applications of the multicanonical method, see
\cite{bergReview}). This method uses a Markov process with an
extended Metropolis accepting ratio to sample any state $x$ in the
phase space with a probability $P(x)$ proportional to the inverse
of the density of sates $g(E)$,
\begin{eqnarray}
    P(x)\propto \frac{1}{g[E(x)]}  \quad ,
  \label{mmprob}
\end{eqnarray}
with $E(x)$ the energy of state $x$. Then, the Metropolis
accepting ratio for a move from $x$ to $x'$ becomes
\begin{eqnarray}
    A(x'|x) = \min \{1,g[E(x)]/g[E(x')]\} \quad .
    \label{Axx}
\end{eqnarray}
By this method, the histogram of samples on the energy axis,
$V(E)$, is flat, so long $g(E)$ is the actual density of states.
If not, a better approximation $g_1(E)$ can be found by dividing
$V(E)$ into $P(x)$, i.e.,
\begin{eqnarray}
    g_1(x)\propto V(E) g(E)  \quad .
    \label{mmprob1}
\end{eqnarray}
The multicanonical method is actually an iterative procedure that,
first, proposes some trial function $g(E)$ and, then, improves it
by successive applications of the procedure described above.

As every Markov process does, the multicanonical method needs a
number $t_{\rm eq}$ of steps before reaching the desired
probability distribution, that is, before starting to take the
samples. For some sampling distributions, like the Boltzmann one,
it is easy to find narrowed-distributed quantities, like the
magnetization, such that $t_{\rm eq}$ can be measured just by
tracking the process until that quantity oscillates around a mean
value and inside a 10\% confidence interval, for instance. That is
what is called {\it statistical control}. The well-established
procedure of computing the non-linear correlation function and the
non-linear correlation time \cite{Binder} is just a sophistication
of this procedure, which gives good results if the quantity
evolves to its equilibrium distribution as an exponential decay.
For the multicanonical method, in contrast, it is not so easy to
imagine a narrow-peaked quantity that can be used for this
purpose. The typical procedure to find $t_{\rm eq}$ is to run two
or more instances of the system, and to keep track in all of them
of some quantity of interest, i.e. the magnetization, until all
instances join together. The time to equilibrium $t_{\rm eq}$ is
taken as the number of steps, measured in Monte Carlo steps per
site (MCSS), before this joint. Despite the fact that this second
criterion is easy to improve, it deserves the statistical
precision of the first one.

This paper introduces a new criterion to measure $t_{\rm eq}$ and
uses it to investigate the time to equilibrium of the
multicanonical sampling method. The new criterion is based on the
chi-squared ($\chi^2$) deviation between the desired limit
distribution and the actual energy distribution after $t$ steps.
First, we introduce this chi-squared procedure, and we test it
against the more traditional one, based on the magnetization, for
the usual Metropolis sampling with Boltzmann distribution. Second,
we use the chi-squared procedure to investigate the time to
equilibrium for the multicanonical method on 2D-Ising systems, and
how it depends on the system size and the energy. Finally, we
summarize our main results and conclusions.

\section{The Procedure}
%
Consider a sampling method based on a Markov process that takes
any state $x$ with some probability $P(x)$, which is function of
$E(x)$ alone. Starting from some initial distribution $P_0(x)$,
the Markov process transforms it at every step until it reaches
the desired distribution $P(x)$. Let us run $n$ times the Markov
process, starting form the same initial distribution $P_0(x)$ but
running with different seeds of the random number generator. Look
at the energy the system has after $t$ steps on each run, and
cumulate the histogram $V_t(E)$ from these energy values. If
$V_t(E)$ actually corresponds to the limit distribution $P(x)$, it
should approximate the expected probability distribution for $E$,
$P(E)$$=$$g(E)P(x)$. The {\it $\chi^2$ deviation} is a good
statistics to verify this point. It is given by
\begin{eqnarray}
  \label{Chi2Deviation}
  \chi^2(t) = \sum_E {[V_t(E)-n P(E)]^2 \over n P(E)} \quad ,
\end{eqnarray}
where the sum extends over all energy values with more than five
or ten samples. For large $n$ values, $\chi^2$ obeys a {\it
chi-square distribution} of $r$$=$$k_E-1$ degrees of freedom, with
probability density
\begin{eqnarray}
  P(\chi^2)={(\chi^2)^{(r-2)/2} e^{-\chi^2/2}
    \over
    \Gamma(r/2) 2^{r/2}} \quad ,
\end{eqnarray}
where $k_E$ is the number of energy levels in the sum
(\cite{Chi2Deviation}. This distribution has population mean
$\mu$$=$$r$ and standard deviation $\sigma$$=$$\sqrt{2r}$. If we
repeat $m$ times the whole study and look at the average
$\overline{\chi^2}$ of that $m$ values, $\overline{\chi^2}$
distributes like a Gaussian, with $\mu$$=$$r$ and
$\sigma$$=$$\sqrt{2r/m}$. That is, 95,45\% of the values of
$\overline{\chi^2}$ lies between $r-2\sqrt{2r/m}$ and
$r-2\sqrt{2r/m}$. This is only true if $V_t(E)$ corresponds to the
limit distribution $P(E)$. That is what is called {\it statistical
control} \cite{estadistica}.

\begin{figure}[h!]
    \begin{center}
    \includegraphics[scale=0.3]{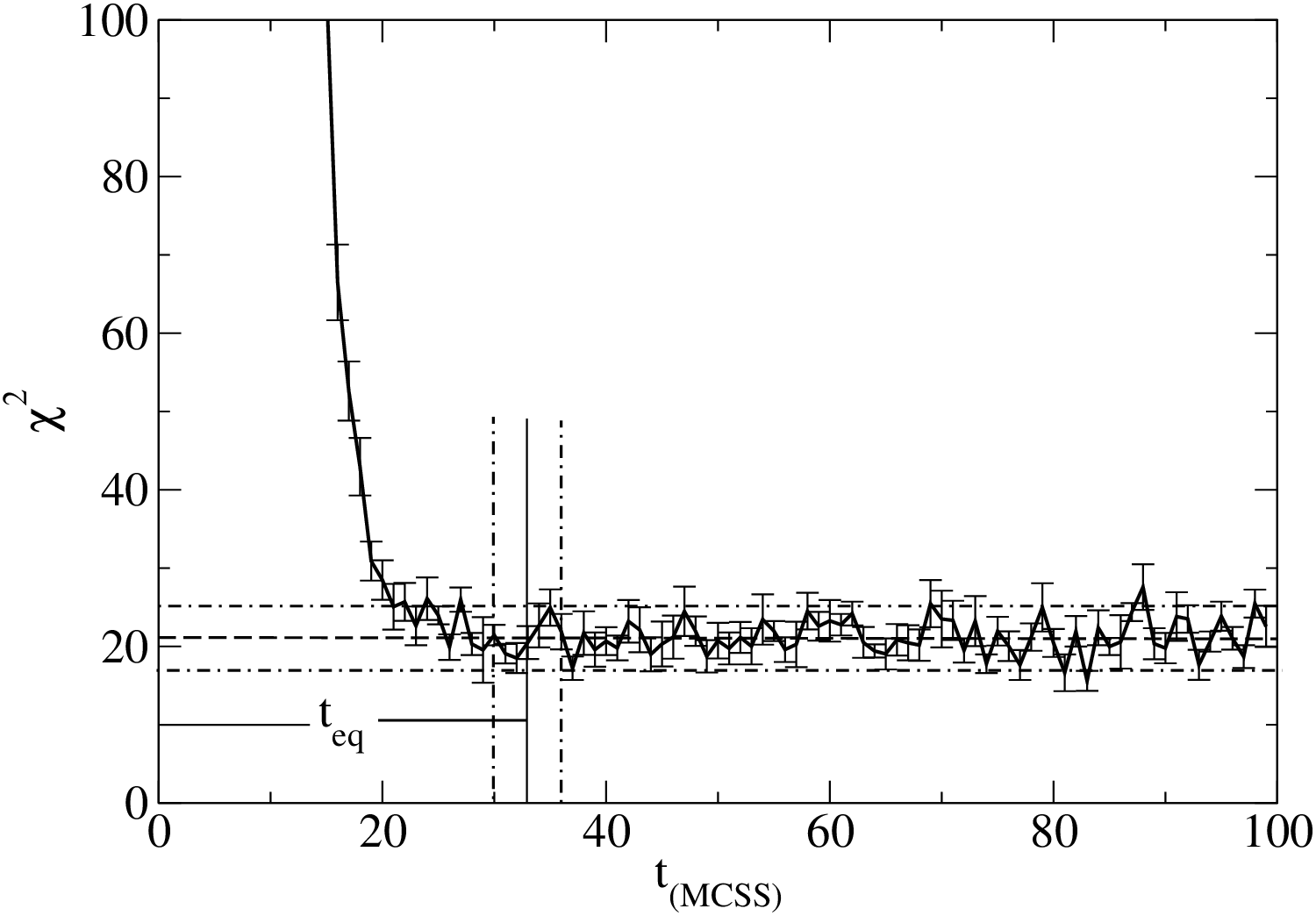}
    \fcaption{\footnotesize
      Example of the  $\chi^2$-criterion to compute the time to
equilibrium
      $t_{eq}$ for a Metropolis sampling method with limit Boltzmann
      distribution ($T$$=$$3.0$) on a $8$$\times$$8$-Ising model.}
    \label{TeqExample}
    \end{center}
\end{figure}

So, the new procedure to find the time to equilibrium is just to
compute $\overline{\chi^2}$, as described above, and to look at
the time, $t_{eq}$, when the statistical control starts. Figure
\ref{TeqExample} shows how this criterion works for a Metropolis
sampling with limit Boltzmann distribution on a
$8$$\times$$8$-Ising model. The temperature was $T$$=$$3.0$, and
the system always started from the state of minimal energy. In
this example, only $k_E$$=$$22$ energy levels were taken into
account to compute $\chi^2(t)$ (Eq. (\ref{Chi2Deviation})), and
the whole study was done $m$$=$$10$ times to compute
$\overline{\chi^2}$, with $n$$=$$10^4$ runs each. Therefore,
statistical control is reached when $\overline{\chi^2}$ oscillates
around $r$$=$$21$, and between $r-2\sqrt{2r/m}$$=$$16.9$ and
$r+\sqrt{2r/m}$$=$$25.1$. We found $t_{eq}=33(3)$.

\begin{figure}[h!]
    \begin{center}
    \includegraphics[scale=0.4]{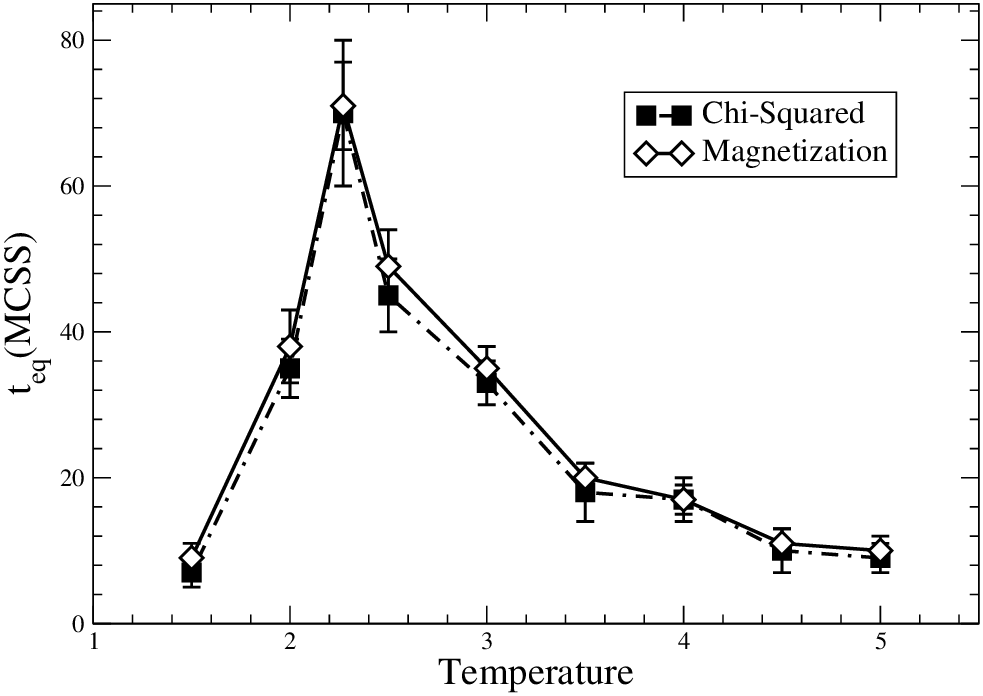}
    \fcaption{\footnotesize
      Time to equilibrium
      $t_{eq}$ for a Metropolis sampling method with limit Boltzmann
      distribution at several temperatures on a $8$$\times$$8$-Ising
model,
      computed both with the new $\chi^2$-criterion and by looking at
the
      statistical control on the magnetization. The two criteria agree
for all
      sampled temperatures.}
    \label{Chi2yMagVsT}
    \end{center}
\end{figure}

Figure \ref{Chi2yMagVsT} shows $t_{eq}$ for the same
$8$$\times$$8$-Ising model, when sampled by the Metropolis method
with limit Boltzmann distribution at several temperatures. Two
criteria were used to measure $t_{eq}$: the $\chi^2$-criterion
described above and another one based on the statistical control
on the magnetization $M$, as follows: we looked at the
magnetization the system had after $t$ steps on each one of the
$n$ runs and computed their mean value $\lb M(t) \rb$. The time to
equilibrium was chosen as the time when statistical control
started, with $\mu$$=$$\lb M \rb_T$ (the average magnetization at
temperature $T$) and $\sigma$$=$$\sqrt{T \chi_{\rm s}(T)/m}$ (with
$\chi_{\rm s}(T)$ the magnetic susceptibility at temperature $T$,
$\chi_{\rm s}(T)$$=$$1/k_{\rm B}T [\lb M^2 \rb_T - \lb M
\rb_T^2]$). All other parameters in the simulation were the same
as in figure \ref{TeqExample}. We can see an excellent agreement
between both criteria for all temperatures. That is, our new
criterion works well by measuring the time to equilibrium of this
Markov process. \vspace{0.8cm}

\section{The multicanonical case}
%
With the new $\chi^2$-criterion on hand, we investigate the time
to equilibrium for the multicanonical sampling method on
two-dimensional Ising models of several sizes. For all simulations
the actual density of states $g(E)$, computed by P.D. Beale
\cite{beale}, was used from start to perform every multicanonical
step. In all cases we chose $n$$=$$10^4$ and $m$$=$$10$, as
before.

\begin{figure}[h!]
\vspace{0.8cm}
    \begin{center}
    \includegraphics[scale=0.3]{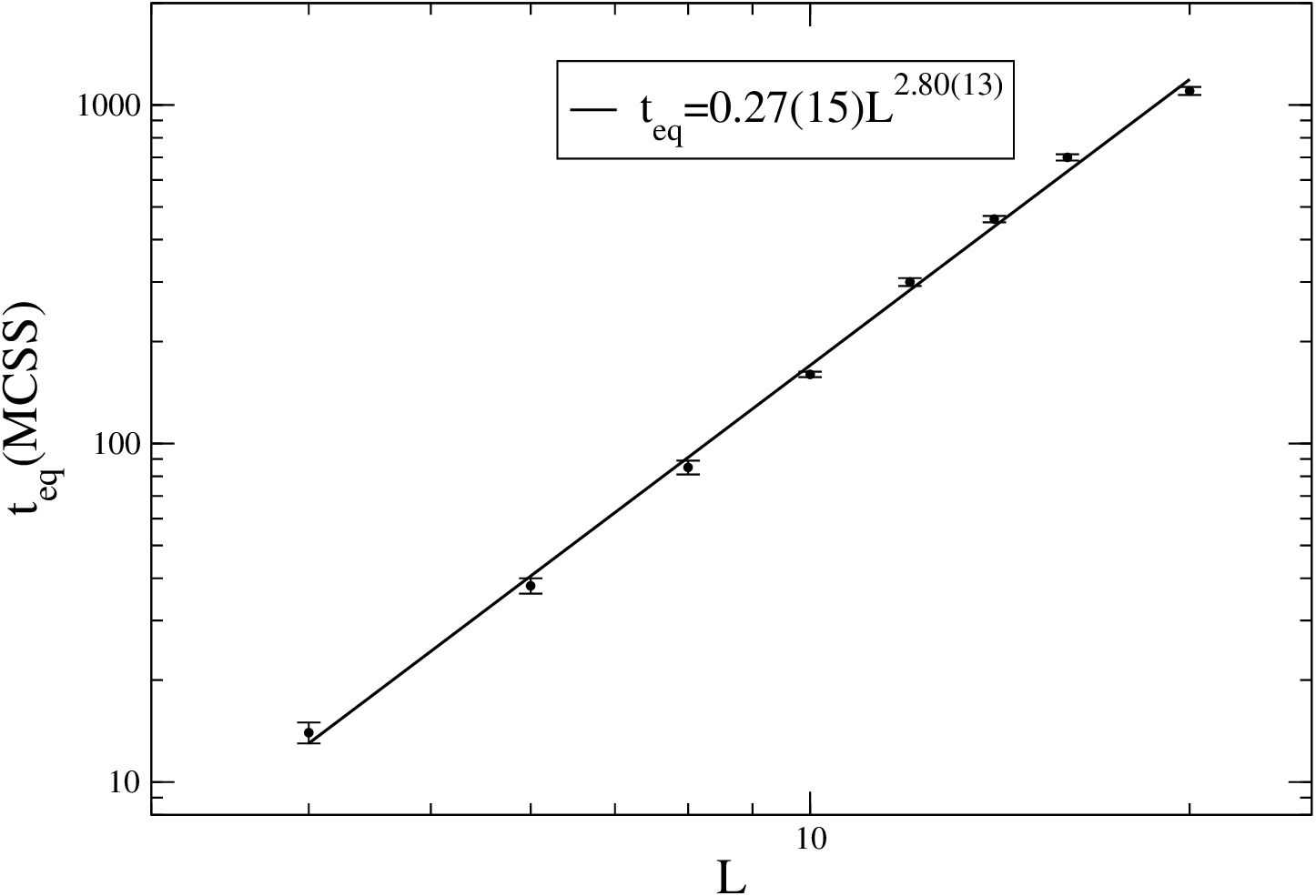}
    \fcaption{\footnotesize
      Dependence of the time to equilibrium
      $t_{eq}$ on system size $L$ for the multicanonical sampling
method on
      $L$$\times$$L$-Ising models. The true density of states $g(E)$
was used
      from start at every multicanonical step.}
    \label{teqVsL}
    \end{center}
\end{figure}

Figure \ref{teqVsL} shows time to equilibrium against system size,
when the whole energy range of positive temperatures
$-2L^2$$\le$$E$$\le$$0$ is sampled. Through linear regression, we
obtain a power-law behavior, $t_{\rm eq}$$=$$0.27(15)L^{2.80(13)}$
\footnote{All error bars in this paper are two-sigma.}, with
correlation coefficcients $r$$=$$0.9988$ and $F$$=$$2416.72$. That
is, the time to equilibrium increases with system size like
$L^{2.80(13)}$.

\begin{figure}[h!]
  \begin{center}
    \includegraphics[scale=0.3]{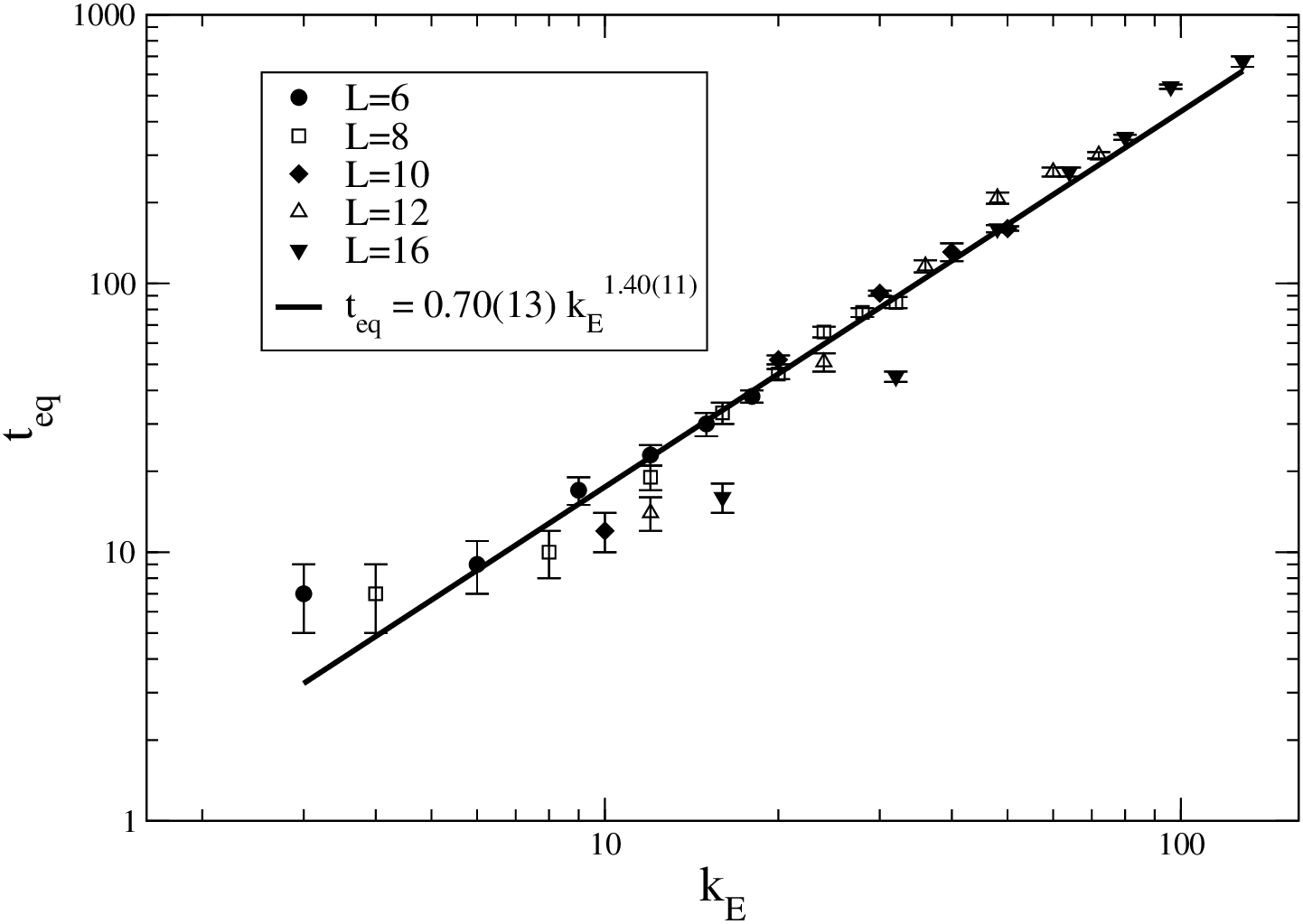}
    \fcaption{\footnotesize Time to equilibrium $t_{eq}$ against the
number $k_E$ of energy
      levels sampled by the multicanonical method, on two-dimensional
Ising
      models of system sizes $L$$=$$6,\,8,\,10,\,12$ and $16$. All
energy
      intervals to be sampled start at $E_{inf}$$=$$-2L^2$ and contain
$k_E$
      energy levels.}
    \label{teqVskE}
  \end{center}
\end{figure}

Next, we look at how the time to equilibrium changes with the
number of energy levels (the number of bins in the histogram)
sampled by the multicanonical method. For that purpose a set of
multicanonical simulations was performed on Ising models of system
sizes $L$$=$$6,\,8,\,10,\,12$ and $16$. All energy intervals start
from $E_{inf}$$=$$-2L^2$ and extend up through $k_E$ energy
levels. Figure \ref{teqVskE} shows $t_{eq}$ against $k_E$ for all
system sizes. We obtain, again, a power-law behavior,
$t_{eq}$$=$$0.70(13)\, k_E^{1.40(11)}$, with correlation
coefficients $r$$=$$0.9783$ and $F$$=$$598.19$. Summarizing,
$t_{eq}$ grows like $k_E^{1.40(11)}$.

Finally, we hold constant the relative length of the energy
interval and we investigate how the time to equilibrium changes
with the position of this interval along the energy axis. With
$E_{min}$$=$$-2L^2$, an energy interval of length $0.16 |E_{min}|$
was set at different places on the energy axis, actually around
approximate energy values of $0.92$, $0.84$, $0.75$, $0.68$,
$0.59$, $0.52$, $0.43$, $0.36$, $0.25$ and $0.10$ times $E_{min}$.
The system reached the energy interval from below by starting at a
configuration of minimal energy and by accepting every random flip
that increases the energy. This procedure fix the initial
distribution $P_0(x)$. 
A whole set of simulations was performed on systems of sizes
$L$$=$$6,\,8,\,10,\,12$ and $16$, as before. \vspace{0.8cm}
\begin{figure}[h!]
  \begin{center}
    \includegraphics[scale=0.3]{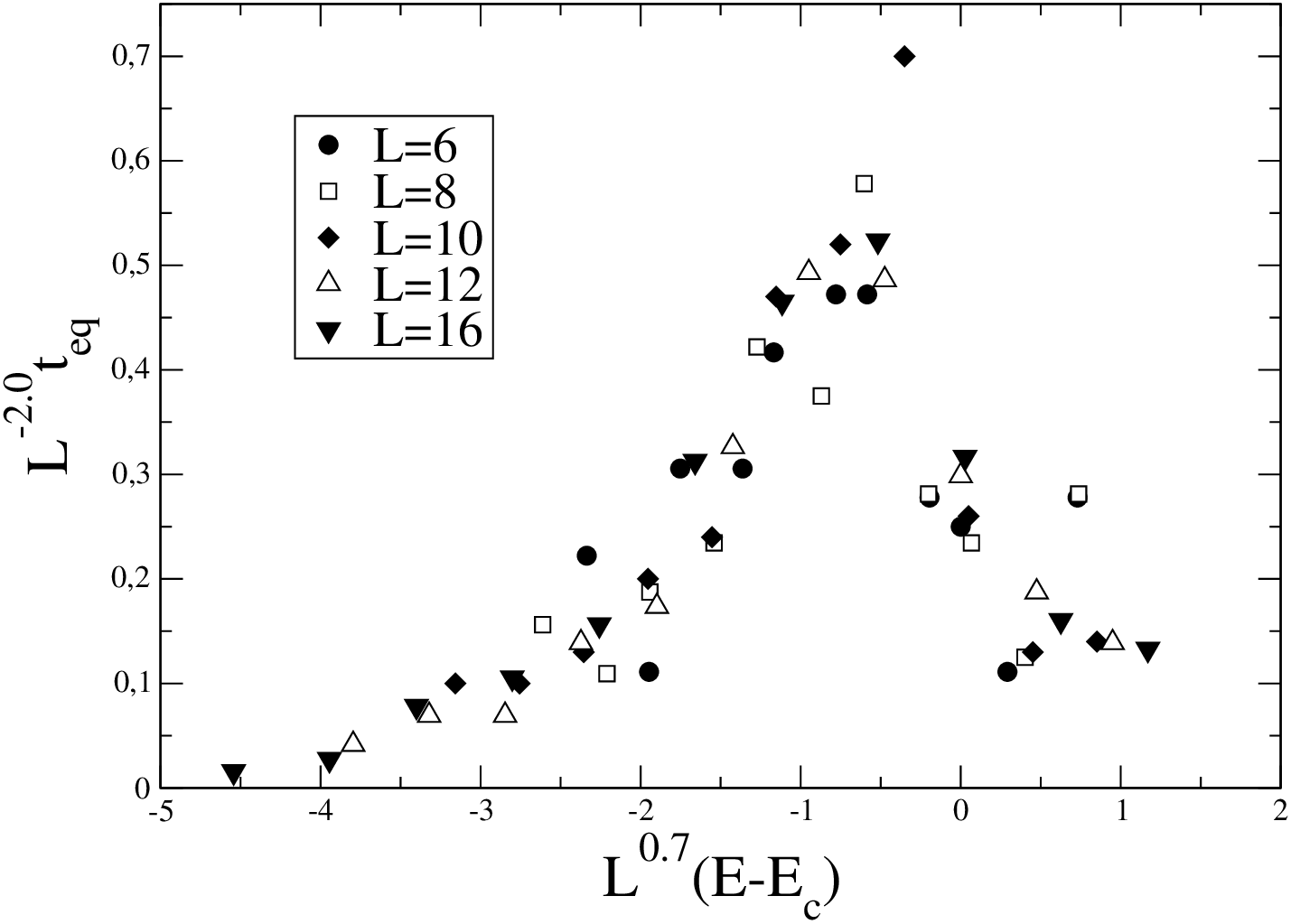}
    \fcaption{\footnotesize Time to equilibrium $t_{eq}$ for the
multicanonical samplig
      method on energy intervals of length $0.16 |E_{min}|$
      (with $E_{min}$$=$$-2L^2$), placed at different places on the
      energy axis. The horizontal axis represents the value for the
central
      energy $E$ in the interval. A certain kind of critical slowing
down can
      be observed around a critical energy of $E_{\rm c}$$=$$0.7540328
      E_{min}$. The simulations were performed on two-dimensional Ising
      systems of sizes $L$$=$$6,\,8,\,10,\,12$ and $16$.}
    \label{SlowDown}
  \end{center}
\end{figure}
Figure \ref{SlowDown} summarizes these results. We try a
finite-size scaling around the critical energy $E_{\rm
c}$$=$$0.7540328 E_{min}$ (the mean energy at the critical
temperature $T_{\rm c}$$\simeq$$2.269185$ for an infinite-size
system) by plotting  $L^{-2.0} t_{eq}$ against $L^{0.7} (E-E_{\rm
c})/E_{\rm c}$. Although the scaling is not a perfect one, we
observe a certain kind of {\it critical slowing down}, i.e. the
time to equilibrium $t_{eq}$ grows if the energy interval lies
near the critical energy and diverges with increasing system size.

\section{Conclusions}
%
As every Markov process does, multicanonical sampling needs some
steps before reaching the desired sampling distribution. We have
found that this time to equilibrium (measured in MCSS) changes
mainly with the number of energy bins, $k_E$, to be sampled. Our
main result is that this dependence obeys a power law,
$t_{eq}$$=$$0.70(13)\, k_E^{1.40(11)}$, at least for the
two-dimensional Ising model. If one wants to sample the whole
energy interval of positive temperature for several system sizes
(which consists of $k_E$$=$$L^2/2$ energy bins), this result
predicts that $t_{eq}$ will grow with system size, $L$, as
$t_{eq}$$=$$0.27(5)\, L^{2.80(22)}$. A direct measurement gave us
$t_{\rm
  eq}$$=$$0.27(15)L^{2.80(13)}$, in perfect agreement with the
prediction above.

This result has important consequences. First, it supports the
usual procedure with multicanonical samplings of dividing the
energy interval of interest in many small intervals and sampling
each one separately. This is not just to parallelize the method,
it actually reduces the total time to equilibrium, because it
grows faster than linear with $k_E$. Second, it says us that we
have to be much more careful by performing multicanonical
samplings on more than one parameter. If we construct an histogram
on two parameters (let us say, the energy and the magnetization,
for instance) with $k$ bins on each one, the time to equilibrium
will grow like $k^{2.80}$, and that can be much faster than
expected (so far our results can be extended to this - somewhat
different - situation).

By dividing the energy range of interest in a fixed number of
small intervals and by using a multicanonical sampling on each
one, we have found that the equilibration times is larger for
energy intervals around the critical energy (the mean energy at
critical temperature), and that those times grow with system size
like $L^{2.0}$, approximately. Although the finite-size scaling is
not perfect, it tells us that we have to be more careful on energy
intervals around the critical energy, and that the time to
equilibrium should be measured just there, if we want to assign
the same number of MCSS equilibration steps for all intervals.

All these results were obtained by using a new criterion to
estimate the time to equilibrium of any sampling Markov process.
We have shown that this new criterion gives the same equilibration
times than the usual one (based on the magnetization) for the
canonical Metropolis sampling of a 2D-Ising model at several
temperatures. Although general, this new method is especially
useful for sampling procedures that construct a flat histogram on
the energy axis, like the multicanonical method, the
flat-histogram method of J.-S. Wang \cite{wang}, the sampling
method of F. Wang and D. P. Landau \cite{landau} or the
broad-histogram sampling of P.M.C. de Oliveira \cite{ancho}. We
are sure that our criterion will work properly for all those
sampling procedures, revealing important issues of its behavior
that can encourage its proper use and improve its performance.

\nonumsection{References}


\begin{thebibliography}{000}
\bibitem{berg}
B.\ A.\ Berg, {\bibit J.\ Mod.\ Phys.} {\bibbf C3}, 1083
(1992).

\bibitem{berg0}
B.\ A.\ Berg, {\bibit Comp.\ Phys. Comm.} {\bibbf 153}, 397
(2003).

\bibitem{lee}
J.\ Lee, {\bibit Phys.\ Rev.\ Lett.} {\bibbf 71}, 211 (1993).

\bibitem{berg1}
B.\ A.\ Berg and T.\ Neuhaus, {\bibit Phys.\ Rev.\ Lett.} {\bibbf
B267}, 249 (1991).

\bibitem{berg2}
B.\ A.\ Berg and T.\ Neuhaus, {\bibit Phys.\ Rev.\ Lett.} {\bibbf
68 }, 9 (1992).

\bibitem{bergBilloireJanke2002}
B.\ A.\ Berg, A.\ Billoire and W.\ Janke, {\bibit Phys.\ Rev.}
{\bibbf E65}, 045102R (2002).

\bibitem{LimaMenezes}
A.\ R.\ Lima and M.\ Argollo de Menezes, {\bibit Phys.\ Rev}
{\bibbf E63}, 020106R (2001).

\bibitem{PengHansmann2003} Y.\ Peng, U.\ H.\ E.\ Hansmann and N.\ A.\ Alves, {\bibit J.\ Chem.\ Phys.}
{\bibbf 117}, 2374 (2003).

\bibitem{ArkinCelik} H.\ Arkin and T.\ Celik, {\bibit Int.\ J.\ of Mod.\
Phys.} {\bibbf C14}, 113 (2002).

\bibitem{HansmannOkamoto}
U.\ H.\ E.\ Hansmann and Y.\ A.\ Okamoto, {\bibit J.\ Comput.\
Chem.} {\bibbf 14}, 1333 (1993).

\bibitem{ReichertGahler}
M.\ Reichert and F.\ Gahler, {\bibit cond-mat/0302070}, (2003).

\bibitem{bergReview}
B.\ A.\ Berg, {\bibit Fields Inst.\ Commun.} {\bibbf 26}, 1
(2000).

\bibitem{Binder}
K.\ Binder and D.\ W.\ Heermann, in {\it Monte Carlo Simulation in
Statistical Physics:An Introduction} (Ed.\ Heidelberg Springer,
1997), p.150.

\bibitem{estadistica}
R.\ E.\ Walpole and R.\ H.\ Myers, in {\it Probability and
statistics for engineersand scientists}, (Macmillan Publishing
Co., 1991), p.677.

\bibitem{beale}
P.\ D.\ Beale, {\bibit Phys.\ Rev.\ Lett.} {\bibbf 76}, 78
(1996).
\bibitem{wang}
J.\ -S.\ Wang, {\bibit Physica A} {\bibbf 147}, 281 (2000).

\bibitem{landau}
F.\ Wang and D.\ P.\ Landau, {\bibit Phys.\ Rev.\ Lett.} {\bibbf
85}, 2050 (2001).

\bibitem{ancho}
P.\ M.\ C.\ de Oliveira, T.\ J.\ P.\ Penna and H.\ J.\ Herrmann,
{\bibit Braz.J. of Phys.} {\bibbf 26}, 677 (1996).

\bibitem{Chi2Deviation}
R.\ E.\ Walpole, in {\it Introduction to statistics}, (Macmillan
Publishing Co., 1982), p.275.
\end{thebibliography}
\end{document}